\documentclass[useAMS,usenatbib]{mn2e}
\usepackage{epsfig}

\title[The properties of 70$\mu$m-selected high-redshift galaxies in
  the Extended Groth Strip]{The properties of 70$\mu$m-selected
  high-redshift galaxies in the Extended Groth Strip}
\author[M.
  Symeonidis]{M. Symeonidis$^{1}$\thanks{myrtos@astro.ox.ac.uk},
  S. P. Willner$^{2}$, D. Rigopoulou$^{1}$, J.-S. Huang$^{2}$,
  G. G. Fazio$^{2}$ \and and M. J. Jarvis$^{3}$\\
$^{1}$Department of Astrophysics, University of Oxford, Denys Wilkinson Building, Keble Road, Oxford OX1 3RH, UK\\
$^{2}$Harvard-Smithsonian Center for Astrophysics, 60 Garden street, Cambridge, MA 02138, USA\\
$^{3}$Centre for Astrophysics, Science $\&$ Technology Research Institute, University of Hertfordshire, Hatfield AL10 9AB, UK}

\begin{document}

\date{Accepted 2007 December 21.  Received 2007 December 20; in original form 2007 September 4}

\pagerange{\pageref{firstpage}--\pageref{lastpage}} \pubyear{2007}

\maketitle

\label{firstpage}

\begin{abstract}
We examine the infrared properties of 43 high redshift
(0.1\,$<$\,z\,$<$\,1.2), infrared-luminous galaxies in the Extended Groth
Strip (EGS), selected by a deep 70\,$\mu$m survey with the Multiband
Imaging Photometer on \emph{Spitzer} (MIPS). In addition and with
reference to starburst-type Spectral Energy Distributions (SEDs), we
derive a set of equations for estimating the total infrared luminosity
($L_{IR}$) in the range 8--1000\,$\mu$m using photometry from at least
one MIPS band. 42 out of 43 of our sources' optical/infrared SEDs ($\lambda_{observed}$\,$<$\,160
$\mu$m) are starburst-type, with only one object displaying a
prominent power-law near-infrared continuum. For a quantitative analysis, models
of radiation transfer in dusty media are fit onto the infrared photometry,
revealing that the
majority of galaxies are represented by high extinction, A$_v$\,$>$\,35 and for a
large fraction ($\sim$\,50 per cent) the SED turns over into the
Rayleigh-Jeans regime at wavelengths longward of 90\,$\mu$m. For comparison, we also fit semi-empirical
templates based on \emph{local} galaxy data, however, these underestimate the
far-infrared SED shape by a
factor of at least 2 and in extreme cases up to 10 for the majority
($\sim$\,70 per cent) of the sources. Further investigation of SED
characteristics reveals that the mid-infrared (70/24\,$\mu$m)
continuum slope is decoupled from various
galaxy properties such as the total infrared luminosity and
far-infrared peak, quantified by the L$_{160}$/L$_{70}$ ratio. In
view of these results, we propose that these high-redshift galaxies have different properties to
their local counterparts, in the sense that large
amounts of dust cause heavy obscuration and are responsible for
an additional cold emissive component,
appearing as a \emph{far-infrared excess} in their SEDs.

\end{abstract}


\section{Introduction}
The classes of galaxies known as Luminous, Ultraluminous and
Hyperluminous Infrared Galaxies (LIRGs, ULIRGs and HyLIRGs), have been
a major subject of focus in observational cosmology since initial,
groundbreaking studies such as those of Soifer et al. (1986) and
Sanders et al. (1987). These objects, discovered in abundance by the
Infrared Astronomical Satellite (\emph{IRAS}), were found to possess
an array of extreme properties, including an excessive energy output,
effectively manifested as a large infrared/optical ratio (Sanders et
al. 1989). It is now widely accepted that the combination of a
powerful UV-photon source and the presence of substantial amounts of
dust are responsible for such extreme luminosities of the order of
$10^{10}$--$10^{14}$$L_{\odot}$ (e.g. Sanders $\&$ Mirabel 1996, Genzel et
al. 1998). Determining the nature of the central energy source has
been the subject of numerous subsequent studies, very often
supplemented by multiwavelength data (e.g. Gregorich et al. 1995,
Genzel et al. 1998, Rigopoulou et al. 1999, Klaas et al. 2001, Tacconi
et al. 2002, Alonso-Herrero et al. 2005). These have revealed
populations dominated by powerful starburst activity, but with a
non-negligible Active Galactic Nucleus (AGN) contribution found to
increase with bolometric luminosity (e.g. Veilleux et al. 1997, Tran
et al. 2001, Brand et al. 2006).

The processes responsible for energy production and radiation transfer
in the interstellar medium of such obscured systems are primarily
evaluated by studying galaxy Spectral Energy Distributions (SEDs), sometimes replacing
spectroscopic diagnostics in determining the nature of the central engine.  For infrared-luminous
objects, these processes are greatly influenced by the composition,
temperature and distribution of interstellar dust. Quantifying the
infrared energy budget is, therefore, key in determining the generic
properties of a galaxy and, as it is directly coupled to the central
energy source, important in estimating the integrated cosmic star
formation history. Cases where photometric data is scarce, benefit from
various methods of data extrapolation and interpolation: construction
of infrared SEDs has been a popular approach of modelling emission
from dusty systems, either directly from first principles or
semi-empirically, bridging together contributions from different parts
of the interstellar medium (e.g. Guiderdoni et al. 1998, Calzetti et
al. 2000, Rowan-Robinson 2000). Such models have enabled the
examination of sources with extreme properties and
pronounced infrared luminosity, either by solely considering star formation processes or
including contribution from AGN. In addition, establishing
correlations between monochromatic flux densities or luminosities has
proven extremely advantageous and has conveniently been used as a
first order approximation of the infrared energy budget, as well as a
diagnostic for the physical processes in galaxies, adopted extensively
in evolution studies (e.g. Franceschini et al. 2001).

In this paper we investigate the properties of 43 infrared-luminous
galaxies from one of the first deep \emph{Spitzer} far-IR
surveys. This sub-sample is part of a population of 178 sources
detected at 70\,$\mu$m by the Multiband Imaging Photometer on
\emph{Spitzer} (MIPS) (hereafter the 70\,$\mu$m population), down to a
limiting flux density of 4\,mJy at 5$\sigma$. In Symeonidis et
al. 2007 (hereafter S07), we gave a brief overview of the sources'
properties and derived Star Formation Rate (SFR) estimates. Here, we aim to gain qualitative insight into their
physical properties, primarily with respect to dust, comparing them to similar objects in the local universe. For
this purpose and to quantitatively characterise the sample, we fit our
photometry with three starburst dust models and a set of empirical
local galaxy templates. 

The paper is set out as follows: Section 2 introduces our selection criteria for the
sub-sample, including an overview of infrared colours. Section 3
analyses and compares the model SED templates, associating the quality
of fits to the properties of the sources. We investigate the infrared
energy budget, with reference to SED characteristics such as the
continuum slope and far-IR peak, in section 4. Finally, section 5
focuses on the derivation of a set of equations to calculate the total
infrared luminosity in the range 8--1000\,$\mu$m, with at least one
MIPS band, for starburst-type sources. Our conclusions
and discussion are presented in section 6. In all subsequent calculations, we have employed the following values: $H_o=71$ kms$^{-1}$Mpc$^{-1}$, $\Omega_M=0.3$ and  $\Omega_{\Lambda}=0.7$.

\section{Introducing the Sample}

\subsection{Data sets and Selection Criteria}

Our main data set is a 70\,$\mu$m image of the Extended Groth Strip (EGS) field, taken by MIPS. In this $\sim$0.5 deg$^{2}$ field, 
178 sources were recovered down to a limiting flux density of 4 mJy at
5$\sigma$. We direct the reader to Davis et al. (2007) and S07 for a detailed
description of all data sets acquired through the All-Wavelength
Extended Groth Strip International Survey (AEGIS). In this section, we briefly refer to
the Deep Extragalactic Evolutionary Probe 2
(DEEP2) spectroscopic survey, the \emph{Spitzer} Infrared Array Camera (IRAC)
survey and the MIPS survey.

70 and 90 per cent of the MIPS 70\,$\mu$m area is also covered by the
IRAC survey and MIPS 24 and 160\,$\mu$m images, respectively. Within the
overlap areas of all 8, 24 and 160\,$\mu$m images, we recover 100
per cent of our targets, a
total of 122 sources (hereafter, the \emph{full sample}). Of these, 103 also appear in the remaining three IRAC
bands (3.6, 4.5 and 5.8\,$\mu$m). DEEP2 obtained optical spectra for $\sim$14,000 objects in the EGS
with a limiting magnitude of $R_{AB}=24.1$, using the DEIMOS
spectrograph on the Keck telescopes (Davis et al. 2003). In the
overlapping area ($\sim$75 per cent) between the MIPS 70\,$\mu$m image
and the DEEP2 EGS survey, 43 reliable (quality factor $\ge$3)
spectroscopic redshifts ($z_{spec}$) were retrieved in the range
$0.1<z<1.2$, with a mean of 0.57 and a median of 0.51. For the
remaining objects, some made the photometric cut but were assigned a
poor quality redshift.

\begin{table*}
\begin{minipage}{126mm}
\caption{The redshift sample (43 sources) - Object cardinal numbers
  with the prefix EGS70 referring to the 70$\mu$m EGS survey, coordinates,
  spectroscopic redshifts, R-band magnitudes (AB), measured flux densities
  at 24, 70 and 160\,$\mu$m. The MIPS 24\,$\mu$m calibration
  uncertainty is 10 per cent, whereas for MIPS 70 and 160\,$\mu$m it is 20 and 30 per cent, respectively.}
\begin{tabular}{|c|c|c|c|c|c|c|c|}
ID & ra (J2000)& dec (J2000)& z$_{spec}$ & R mag (AB) &f$_{24}$ &f$_{70}$&f$_{160}$ \\
& h m s& deg min arc&  & &(mJy)  &(mJy)  & (mJy)  \\
\hline 
      EGS70-41 &  14 24 26.1  & +53 37 23.0 &      0.45 &       20.9 &    1.3 &     44.1    &      117.6 \\
      EGS70-51 &  14 24 34.3  & +53 36 05.4 &      0.22 &       18.6 &   0.9 &    6.9 &     29.1 \\
      EGS70-55 &  14 24 28.2  & +53 36 44.6 &      0.67 &       21.6 &   0.6 &     16.9 &     60.0 \\
      EGS70-58 &  14 25 02.8  & +53 31 25.5 &      0.69 &       22.3 &   0.3 &     10.4 &      115.2 \\
      EGS70-67 &  14 24 30.1  & +53 35 42.9 &      0.96 &       22.1 &    1.0 &     11.6 &     85.7 \\
      EGS70-70 &  14 24 41.5  & +53 33 43.2 &      0.78 &       21.3 &   0.46 &     14.6 &      104.5 \\
      EGS70-72 &  14 24 55.0  & +53 30 04.6 &      0.39 &       20.4 &   0.76 &     11.8 &     23.0 \\
      EGS70-76 &  14 24 29.6  & +53 31 21.3 &      0.73 &       22.6 &   0.11 &     10.5 &     10.2 \\
      EGS70-77 &  14 23 43.1  & +53 35 06.1 &      0.42 &       19.2 &    1.4 &     22.3 &      148.8 \\
      EGS70-78 &  14 24 08.2  & +53 31 22.3 &      0.37 &       20.3 &    1.0 &     15.5 &     57.2 \\
      EGS70-82 &  14 23 49.6  & +53 31 35.4 &       1.03 &       23.4 &   0.38 &     21.7 &     64.3 \\
      EGS70-84 &  14 23 48.8  & +53 30 09.0 &      0.77 &       22.0 &   0.99 &     13.9 &     99.3 \\
      EGS70-85 &  14 24 01.2  & +53 27 54.4 &      0.43 &       19.9 &   0.95 &     12.0 &     38.1 \\
      EGS70-88 &  14 23 49.4  & +53 26 29.8 &      0.78 &       21.5 &    1.68 &     40.0 &      111.5 \\
      EGS70-93 &  14 23 16.1  & +53 30 47.4 &       1.2 &       22.6 &   0.69 &     14.1 &      130.5 \\
      EGS70-94 &  14 23 30.9  & +53 28 22.5 &       1.02 &       21.7 &   0.67 &     11.5 &     82.5 \\
      EGS70-98 &  14 23 35.4  & +53 26 12.3 &      0.29 &       20.1 &    1.85 &     14.9 &     48.3 \\
     EGS70-100 &  14 23 31.6  & +53 26 25.1 &      0.3 &       19.3 &    2.5 &     35.9 &     68.7 \\
     EGS70-103 &  14 23 39.6  & +53 22 32.0 &      0.25 &       19.0 &   0.9 &     12.9 &     60.6 \\
     EGS70-110 &  14 23 18.2  & +53 24 53.5 &      0.42 &       21.3 &   0.5 &     30.5 &     48.6 \\
     EGS70-114 &  14 23 01.6  & +53 24 20.7 &      0.66 &       20.8 &    1.1 &     16.6 &     76.8 \\
     EGS70-115 &  14 23 12.7  & +53 22 16.0 &      0.38 &       20.1 &    1.2 &     14.2 &     57.5 \\
     EGS70-120 &  14 23 22.8  & +53 19 29.2 &      0.57 &       21.6 &   0.5 &    8.9 &     25.8 \\
     EGS70-121 &  14 23 00.1  & +53 21 51.9 &      0.78 &       22.1 &   0.7 &    7.7 &     25.2 \\
     EGS70-122 &  14 22 50.0  & +53 22 53.3 &      0.28 &       19.4 &    1.0 &     14.6 &     74.1 \\
     EGS70-124 &  14 22 39.1  & +53 24 20.3 &      0.85 &       21.3 &    1.1 &     13.9 &     61.1 \\
     EGS70-125 &  14 22 47.3  & +53 22 20.2 &      0.25 &       20.7 &   0.7 &     11.3 &     44.5 \\
     EGS70-126 &  14 23 14.7  & +53 17 59.5 &      0.42 &       21.2 &   0.6 &     29.5 &     41.4 \\
     EGS70-134 &  14 22 56.3  & +53 17 10.1 &      0.75 &       21.9 &   0.4 &    4.8 &     46.2 \\
     EGS70-137 &  14 23 05.7  & +53 15 28.1 &      0.48 &       20.1 &   0.75 &     14.4 &      103.5 \\
     EGS70-138 &  14 22 22.3  & +53 21 58.2 &      0.42 &       21.1 &   0.26 &    7.5 &     32.5 \\
     EGS70-140 &  14 22 25.4  & +53 20 30.0 &      0.46 &       19.4 &   0.57 &     10.7 &     43.8 \\
     EGS70-141 &  14 22 11.8  & +53 19 50.0 &      0.70 &       22.7 &   0.54 &     11.4 &    7.3 \\
     EGS70-144 &  14 22 36.8  & +53 15 01.1 &      0.19 &       19.3 &   0.75 &     13.0 &     37.0 \\
     EGS70-146 &  14 22 18.1  & +53 16 43.6 &      0.46 &       20.1 &   0.64 &    8.9 &     28.5 \\
     EGS70-147 &  14 21 54.3  & +53 19 51.4 &      0.67 &       22.3 &   0.49 &    8.8 &     44.2 \\
     EGS70-149 &  14 22 21.1  & +53 14 33.9 &      0.51 &       20.3 &   0.69 &     14. &     64.6 \\
     EGS70-150 &  14 21 55.4  & +53 18 10.4 &      0.74 &       22.0 &   0.71 &    6.1 &     22.1 \\
     EGS70-152 &  14 22 31.1  & +53 12 37.5 &      0.37 &       20.0 &   0.98 &     13.1 &      124.4 \\
     EGS70-155 &  14 22 17.7  & +53 14 32.7 &      0.17 &       18.6 &    1.2 &     16.7 &     11.2 \\
     EGS70-158 &  14 21 37.5  & +53 17 17.8 &      0.93 &       22.1 &    1.4 &     30.2 &      143.4 \\
     EGS70-165 &  14 21 55.9  & +53 14 22.1 &      0.77 &       21.7 &    2.8 &     16.7 &     28.0 \\
     EGS70-172 &  14 21 41.1  & +53 15 03.8 &      0.48 &       23.1 &   0.2 &     13.2 &     85.7 \\
\end{tabular}
\end{minipage}
\end{table*}

\begin{figure}
\epsfig{file=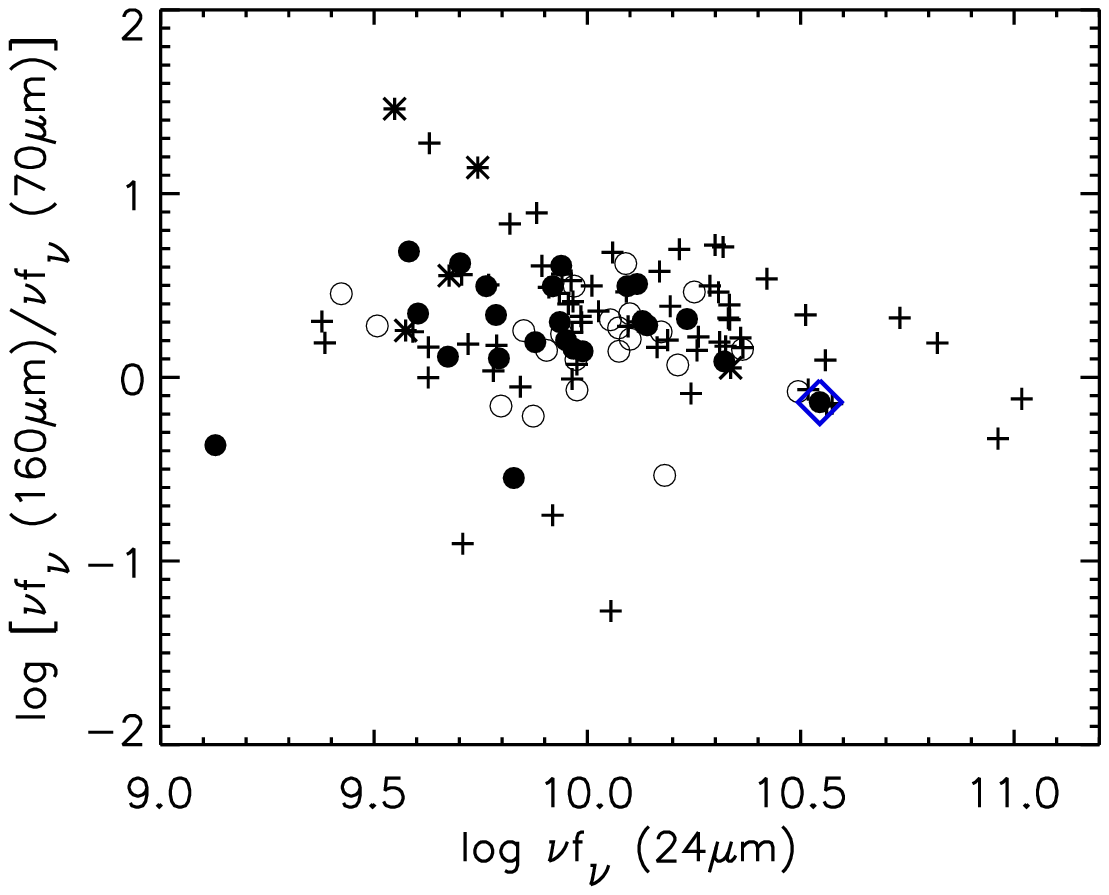,width=8cm}
\caption{Infrared colours for the full sample - ratio of 160\,$\mu$m $\nu f_{\nu}$ to 70\,$\mu$m $\nu
  f_{\nu}$ versus 24\,$\mu$m $\nu f_{\nu}$. Asterisks represent
  R\,$>$\,24.1 sources with no z$_{spec}$, while the rest of the
  symbols are divided into z\,$\leq$\,0.5 (empty circles), z $>$ 0.5 (filled
  circles) and R\,$<$\,24.1 sources with no z$_{spec}$ (crosses). The blue diamond symbol corresponds to the object we identified as AGN-dominated (see discussion in section 2.2). }
\label{fig:mipscolours}
\end{figure}

For the purpose of this work we select a sub-sample of the 70 $\mu$m
population (table 1) (hereafter, the \emph{redshift sample}), which we
require to be within the common area of the IRAC 8\,$\mu$m and the
MIPS 24 and 160\,$\mu$m images, in order to sample the entire
available infrared SED. The criteria are as follows:
\begin{itemize}
\item 8\,$\mu$m photometry from the IRAC survey
\item 24 and 160\,$\mu$m photometry from the MIPS surveys
\item a reliable spectroscopic redshift from DEEP2
\end{itemize}
It is the latter criterion which forms the basis of the
selection, as the former two are always satisfied within the overlap
regions. 34 objects in the redshift sample also appear in the 3.6,
4.5 and 5.8\,$\mu$m IRAC bands. The MIPS (24, 70 and 160\,$\mu$m)
colours of the redshift and full samples are similar (figure
\ref{fig:mipscolours}); apart from a few objects at the
extremes of the 160/70 colour, for the majority of the population the
received flux at 160\,$\mu$m is more than twice that at 70\,$\mu$m and
shows no dependence on redshift or the strength of the 24\,$\mu$m mid-IR emission.

\begin{figure}
\epsfig{file=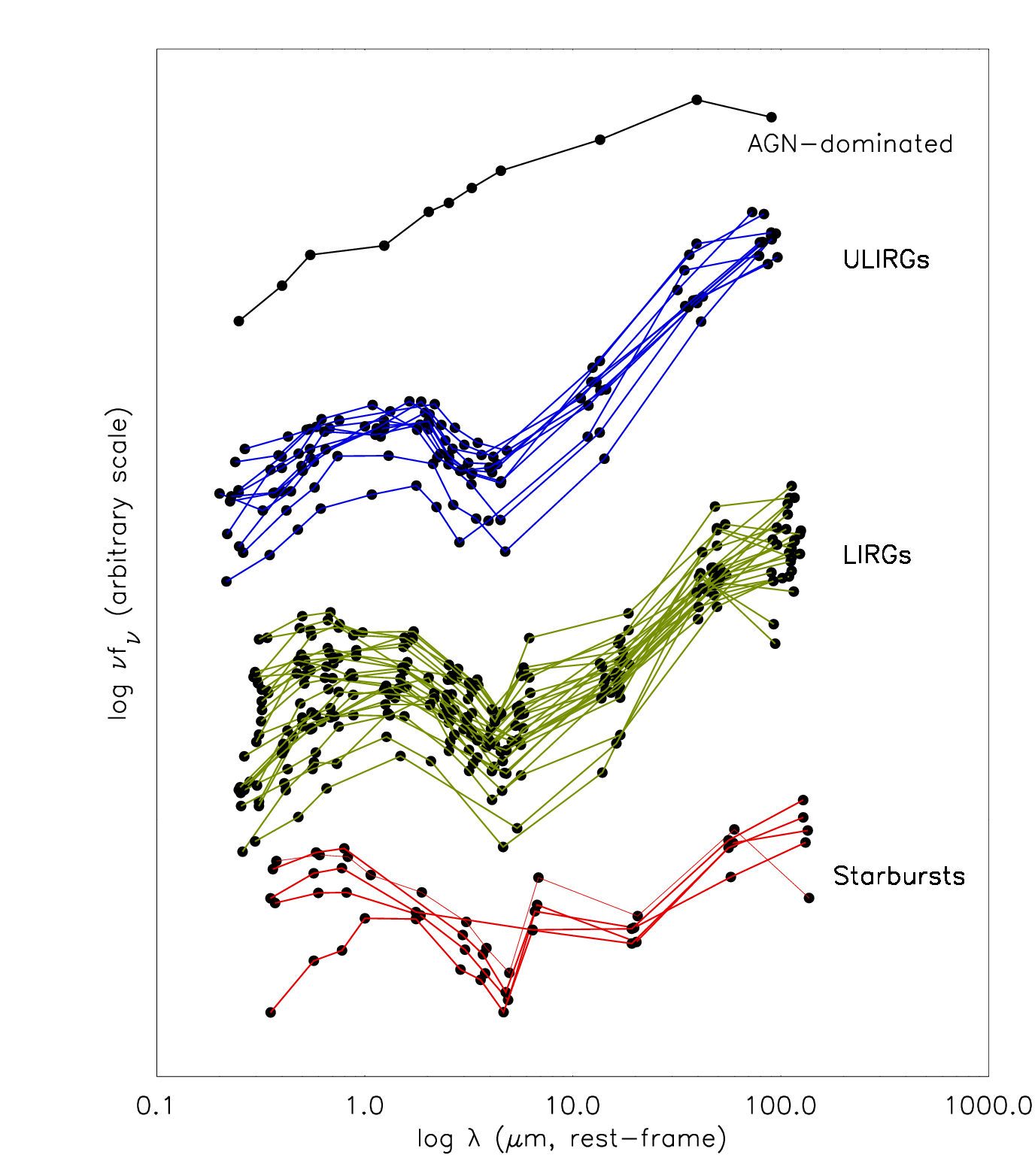,width=8cm}
\caption{The optical to far-IR SEDs of the 43 objects in the redshift
  sample. They are split into 3 luminosity classes - Starbursts (red),
  LIRGs (green) and ULIRGs (blue). Due to the flux density limits of
  the 70\,$\mu$m survey, an increase in total infrared luminosity
  corresponds to increasing redshift, with all starbursts at z $\leq$
  0.25 and all ULIRGs at z\,$\geq$\,0.65. The AGN-dominated SED (see section 2.2) is shown in black. The photometry
  (\textit{BRIJK}, 3.6, 4.5, 5.8, 8, 24, 70 and 160\,$\mu$m) is
  redshift-corrected.}
\label{fig:seds}
\end{figure}

\subsection{Spectral Energy Distributions}
The SEDs of the 43 sources in the redshift sample are shown in figure
\ref{fig:seds}. They are separated into three groups with respect to
the total infrared luminosity in the 8--1000\,$\mu$m range ---
Starbursts ($10^{10}-10^{11} L_{\odot}$), LIRGs ($10^{11}$--$10^{12}$
$L_{\odot}$) and ULIRGs ($10^{12}-10^{13} L_{\odot}$) (see section 4). The SED of an object which displays a clear power-law near-IR
continuum is in black. This was eliminated through examination of IRAC colours
commonly used to identify objects whose SED is AGN-dominated (e.g. Lacy
et al. 2004, Stern et al. 2005).

An obvious trend in the SEDs is the reduction of the stellar bump with
respect to the infrared continuum, for objects of increasing luminosity and, as this is a flux limited survey, redshift. A similar observation was made by Alonso-Herrero et
al. (2006), who found the stellar bump greatly reduced or completely
wiped out in local ULIRGs. This could be a direct consequence of an
increasing AGN versus starburst dominance, indicating
the tendency towards a power-law near-IR continuum or simply due to
the fact that the more
infrared-luminous galaxies will be
fainter in the optical because of increased extinction. For the lower-redshift objects, the 6.2 and/or 7.7\,$\mu$m PAH
features are redshifted into the  8\,$\mu$m band, which displays a
clear increase in flux. Contrary to the pronounced differences in the near-IR, the mid/far-IR
(24-160\,$\mu$m) part of the SEDs makes them indistinguishable; this
has also been noted by other authors, e.g. Klaas et al. (2001) and it is
potentially indicative of the fact that far-IR emission is independent from the nature of
the central energy source.

\section{SED fitting}

In past years, there have been numerous efforts to characterise and
reproduce galaxy SEDs with model templates (e.g. Rowan-Robinson 1980,
1995, Appleton et al. 2004). These have been developed either
semi-empirically or by applying detailed radiation transfer
calculations; in either case there has been reasonable success
with respect to the models' representation of real systems
(e.g. Efstathiou $\&$ Rowan-Robinson 1990, Granato 1994, Silva 1998,
Rowan-Robinson et al. 2005, Efstathiou, Rowan-Robinson $\&$
Siebenmorgen 2003, Siebenmorgen $\&$ Kr\"ugel 2007). Our motive for SED
fitting is to determine the infrared energy
budget of each source and gain insight into its physical
properties. Since the redshift sample mostly displays SEDs typical of
starburst galaxies, we
require models of radiative transfer through
dusty media, considering solely starburst processes. Accordingly,
object 165, whose SED we evaluate as AGN-dominated (see section 2.2), is not included in our
analysis. Nevertheless, we do not exclude a strong AGN contribution in the remaining galaxies. Simply, at this stage, due to the
extremely obscured nature of these objects, we cannot reach any further
conclusions without exploring their multiwavelength properties, which
is beyond the scope of this paper.

We use the Chary $\&$ Elbaz (2001, hereafter CE01), Dale $\&$ Helou
(2002, hereafter DH02) and Siebenmorgen $\&$ Kr\"ugel (2007, hereafter
SK07) libraries, selected on account of their diverse frameworks
and the fact that they span a wide range in infrared luminosity ($10^{9}$ --$10^{14}
L_{\odot}$), from star-forming galaxies to HyLIRGs. In
subsection 3.1 we touch upon the main characteristics of the models
that are relevant to our study. For a more detailed report, we refer
the reader to the relevant papers.

\begin{figure*}
\epsfig{file=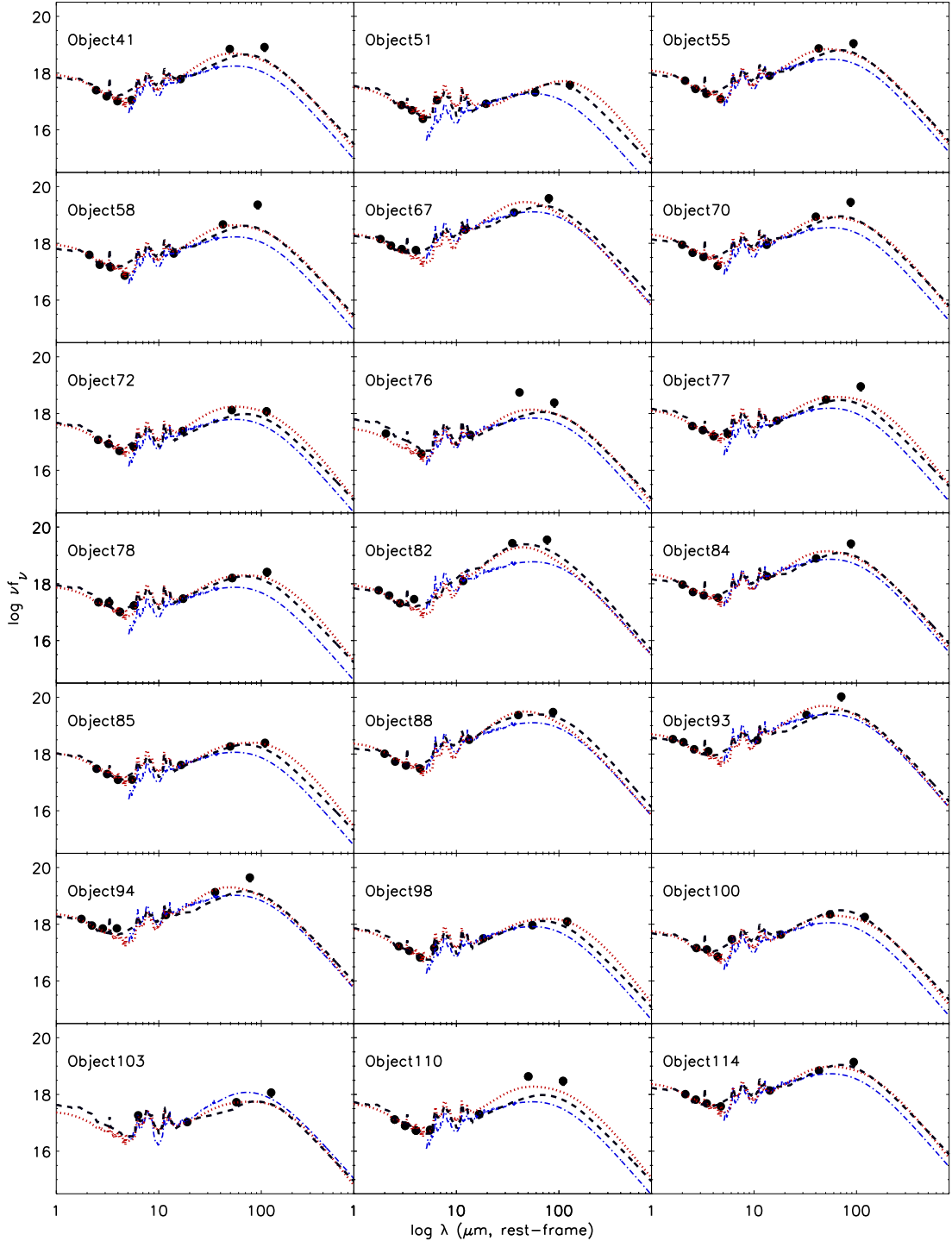,width=15cm}
\caption{SEDs for the first 21 objects from the redshift sample fitted
  with the CE01 (black dashed curve), DH02 (red dotted curve) and
  our empirical templates (blue dot-dashed curve). Only the best-matched (lowest $\chi^2$)
  template is shown. The filled circles are Spitzer
broadband data for the redshift sample, including calibration errors of 10 per cent
for the IRAC and MIPS 24\,$\mu$m photometry, 20 per cent for MIPS
70\,$\mu$m and 30 per cent for MIPS 160\,$\mu$m. All photometry has
  been colour-corrected and redshift-corrected so the plots are at z$\sim$0.} 
\label{fig:badsed}
\end{figure*}

\begin{figure*}
\epsfig{file=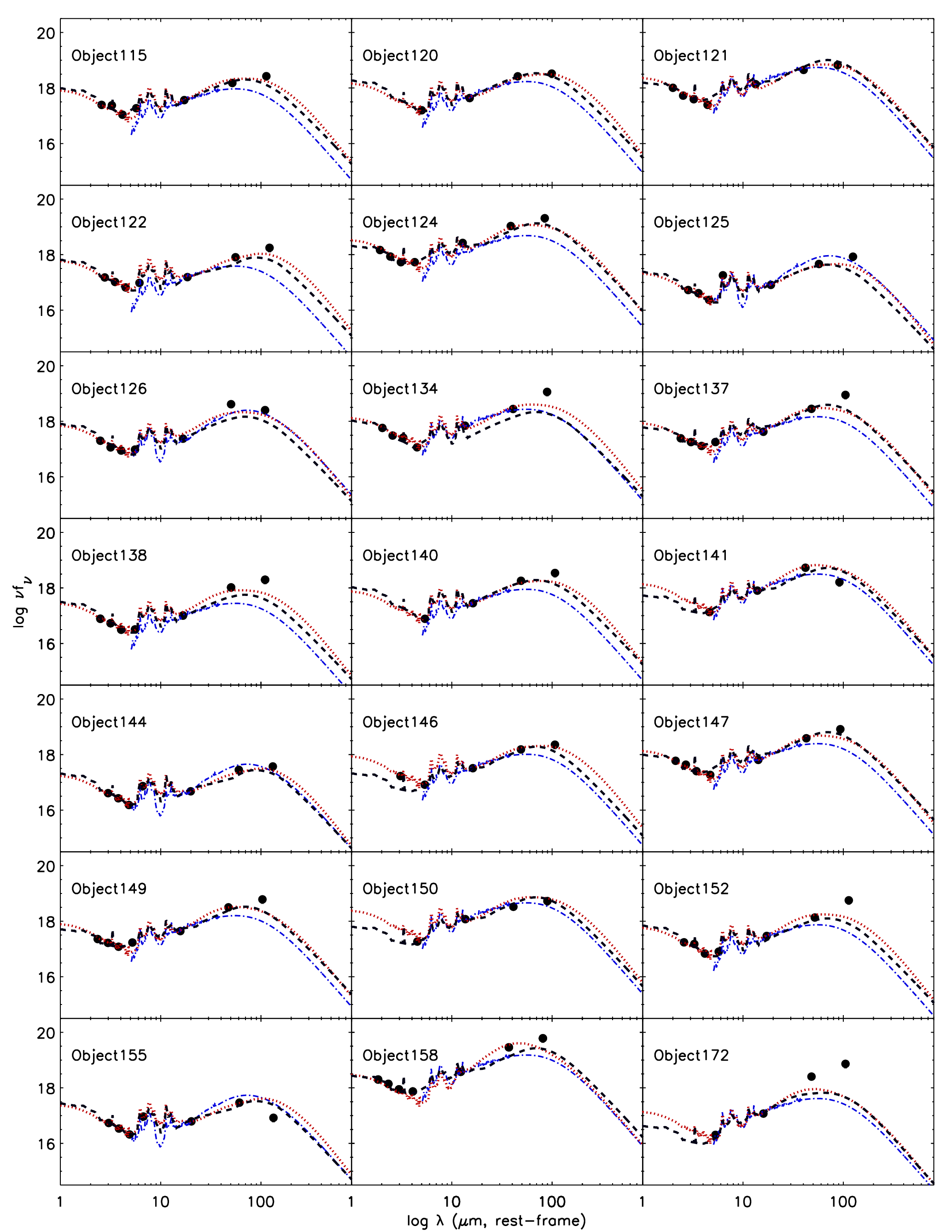,width=15cm}
\caption{SEDs for the remaining 21 objects from the redshift sample
  fitted with the CE01 (black dashed curve), DH02 (red dotted curve)
  and our
  empirical templates (blue dot-dashed curve). Only the
  best-matched (lowest $\chi^2$) template is shown. The filled circles are Spitzer
broadband data for the redshift sample, including calibration errors of 10 per cent
for the IRAC and MIPS 24\,$\mu$m photometry, 20 per cent for MIPS
70\,$\mu$m and 30 per cent for MIPS 160\,$\mu$m. All photometry has
  been colour-corrected and redshift-corrected so the plots are at z$\sim$0.}
\label{fig:badsed2}
\end{figure*}

\begin{figure*}
\epsfig{file=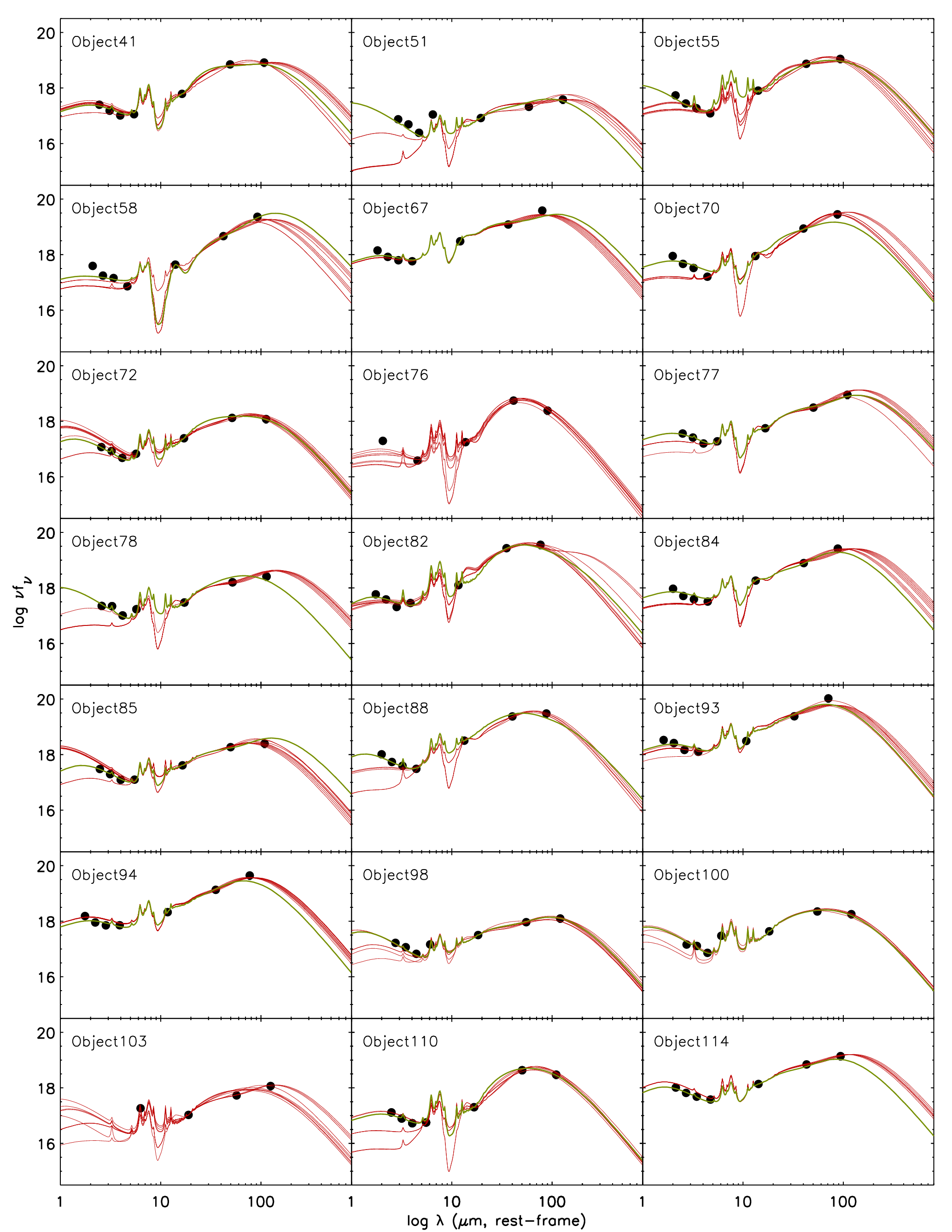,width=15cm}
\caption{SEDs for the first 21 objects from the redshift sample fitted
  with the SK07 templates. The red curves represent the matches with
  the lowest $\chi^2$ values (see section 3.3), fitted on 8, 24, 70
  and 160\,$\mu$m photometry. The green curve represents the match
  with the lowest $\chi^2$ value when fitted on 3.6, 4.5, 5.8, 8, 24,
  70 and 160\,$\mu$m photometry. The filled circles are Spitzer
broadband data for the redshift sample, including calibration errors of 10 per cent
for the IRAC and MIPS 24\,$\mu$m photometry, 20 per cent for MIPS
70\,$\mu$m and 30 per cent for MIPS 160\,$\mu$m. All photometry has
  been colour-corrected and redshift-corrected so the plots are at z$\sim$0.}
\label{fig:goodsed}
\end{figure*}

\begin{figure*}
\epsfig{file=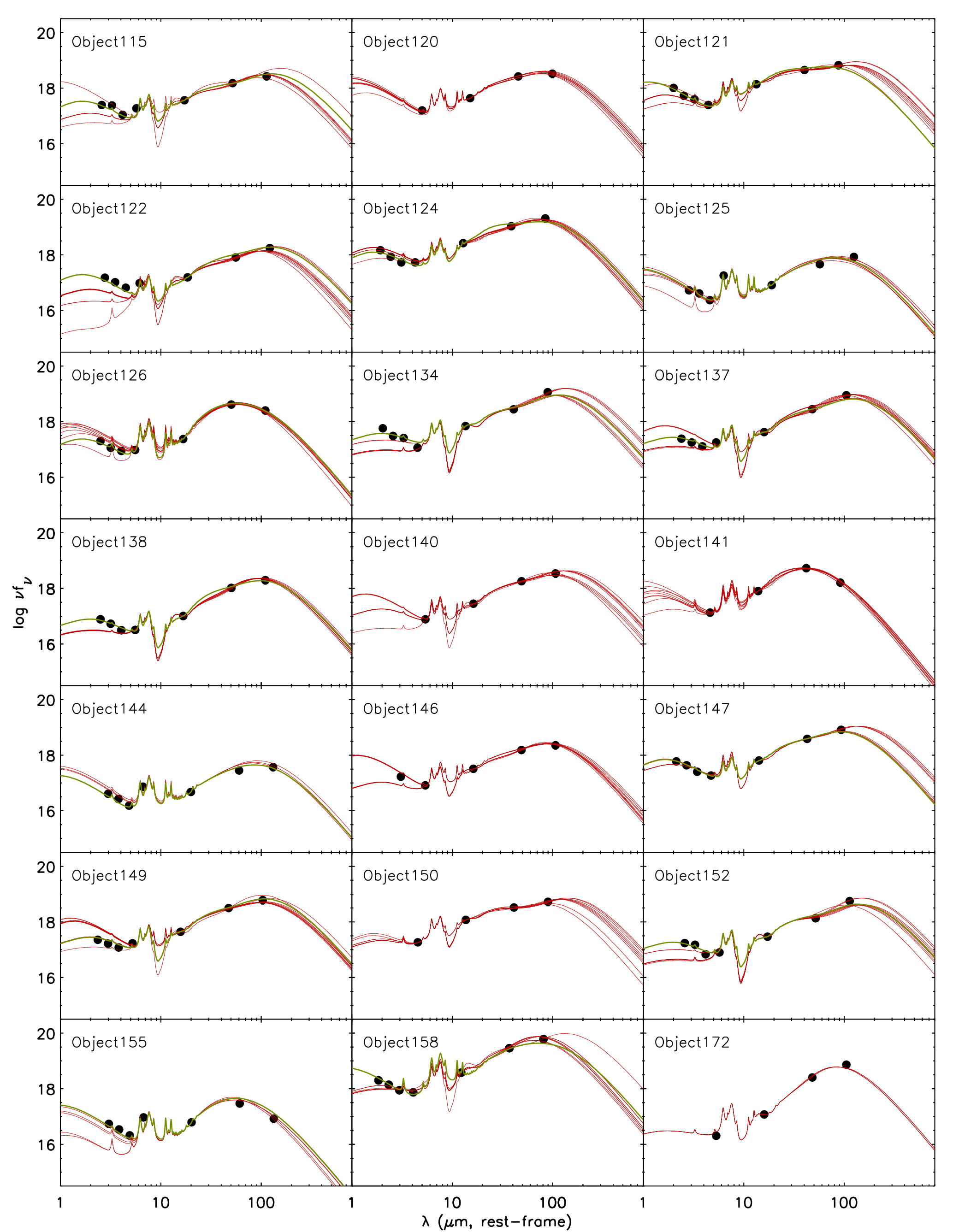,width=15cm}
\caption{SEDs for the remaining 21 objects from the redshift sample
  fitted with the SK07 templates. The red curves represent the matches
  with the lowest $\chi^2$ values (see section 3.3), fitted on 8, 24,
  70 and 160\,$\mu$m photometry. The green curve represents the match
  with the lowest $\chi^2$ value when fitted on 3.6, 4.5, 5.8, 8, 24,
  70 and 160\,$\mu$m photometry. The filled circles are Spitzer
broadband data for the redshift sample, including calibration errors of 10 per cent
for the IRAC and MIPS 24\,$\mu$m photometry, 20 per cent for MIPS
70\,$\mu$m and 30 per cent for MIPS 160\,$\mu$m. All photometry has
  been colour-corrected and redshift-corrected so the plots are at z\,$\sim$\,0.}
\label{fig:goodsed2}
\end{figure*}

\subsection{The dust models}

The CE01 SED templates are based on the models of Silva et al. (1998,
hereafter S98). S98 employ a stellar population synthesis framework,
using Simple Stellar populations (SSP) of different age and
metallicity, including circumstellar dust shells/outflows of AGB
stars that continuously enrich the ISM. Two types of environments are
responsible for the infrared continuum - dust emission associated with
starforming molecular clouds and a cirrus component associated with
the intensity of the stellar field. CE01 generate template SEDs, in
the UV--submm range, of four prototypical galaxies, Arp 220, NGC 6090,
M82, and M51, corresponding to four luminosity classes: ULIRGs, LIRGs,
starbursts and normal galaxies. They partition the four SEDs into mid
and far-infrared components and interpolate between them to create a
set of templates with a range of luminosities, from which they select
the ones which correspond to observed ranges of \emph{IRAS} and
Infrared Space Observatory (\emph{ISO}) colours.

The DH02 templates are based on the work of Desert, Boulanger $\&$
Puget (1990, hereafter DBP90). DBP90 introduce an empirical model
combining emission from Very Large Grains (VLGs) in thermal
equilibrium, stochastically heated Very Small Grains (VSGs) and
Polycyclic Aromatic Hydrocarbon (PAH) features, in radiation fields of
varying intensity. DH02 produce a set of semi-empirical infrared SEDs,
the shape of which is determined by the intensity $U$ of the
interstellar radiation field. They develop a power law dependence of
the dust mass on the value of $U$, the outcome of which is a collection
of dusty environments immersed in a range of radiation fields,
representing a mixture of active and quiescent regions. Their SEDs are
further constrained by \emph{IRAS} and \emph{ISO} observations of
local star-forming galaxies with $L_{IR}$ ranging from  $10^{8}$ to
$10^{12} L_{\odot}$. In addition they vary the emissivity index
$\beta$, which relates the emissivity of dust grains to
the wavelength of emission. 

The SK07 model SEDs are built on the formulation of Kr\"ugel $\&$
Siebenmorgen (1994), who present a modified definition of a starburst
nucleus: instead of a central radiating point source, they employ a
configuration where stars are distributed throughout the volume,
introducing `Hot Spots' - dust regions enveloping OB
stars. Correspondingly, in their calculations, the dust temperature is
varied both in the locality of each star as well as with distance from
the nucleus. Furthermore, since part of the stellar flux is absorbed
within the hot spot and re-radiated in the IR, their formulation replaces some OB stars by
IR-emitting hot spots. This has the effect of reducing the overall
temperature, whilst keeping mid-IR emission unchanged, as this is
considered to be linked to the density of dust within the hot
spot. All other stars are treated separately and are assumed to
contribute to the far-IR part, `cirrus' component of the SED. Their
library of $\sim$\,7000 templates ranges in 5 free parameters, in
physically acceptable combinations: radius of dust emitting region,
total luminosity, visual extinction and hot spot
dust density. The fifth
parameter, $L_{OB}/L_{tot}$, represents the percentage of the total
infrared luminosity that originates from OB stars, compared to the
contribution from the general stellar population and a result it has a direct effect
on the near-IR flux and the representation of the near-IR stellar bump.

\subsection{Our empirical templates}
In addition to the models and as a consistency check, we fit a set of
real galaxy templates, which we construct
by combining spectral and photometric data of local starbursts. The templates are built on a collection of mid-IR (5--38\,$\mu$m)
\emph{Spitzer} InfraRed Spectrograph (IRS) spectra (courtesy of Brandl et al. 2006) from the central
region of 16 local Starbursts, as well as \emph{IRAS} broadband
data (12, 25, 60 and 100\,$\mu$m) from the NASA Extragalactic Database
(NED). The far-IR/submm region (longward of 60\,$\mu$m) is fitted with
a modified black-body emission law, as it is considered to sample
emission from cool dust in equilibrium heated by the interstellar
medium. This
is of the form $B_{\lambda}(T)(1-e^{- \tau _{\lambda}})$, with a
wavelength dependent
optical depth
$\tau_{\lambda}=\tau_{100 \mu m}(100 \mu m/\lambda)^{\beta}$ (e.g. see
Klaas et al. 2001) and a dust emissivity index $\beta$. We assume a low opacity limit, so approximate the
term $(1-e^{- \tau _{\lambda}})$ by
$\lambda^{-\beta}$. Typical reported values of $\beta$ range between
1.5--2 (e.g. Dunne et al. 2000, Lisenfeld, Isaak $\&$ Hills 2000),
so we adopt $\beta$=1.5, consistent with studies of the far-IR emissivity of large
grains (Desert, Boulanger $\&$ Puget 1990). The greybody
temperature is allowed to vary in order to minimise $\chi^{2}$ on the
60--100\,$\mu$m region, resulting in a temperature range of 30--50
K. These 16 empirical SED templates vary with respect to total infrared
luminosity, position of far-IR peak and mid-IR continuum slope.

\begin{table*}
\begin{minipage}{126mm}
\caption{Parameters corresponding to the SK07 templates, fitted with
  photometry at 8, 24, 70 and 160\,$\mu$m, that produced the lowest
  $\chi ^2$ matches (used in subsequent calculations). The columns
  are as follows: Object ID, radius of the central nuclear region
  (kpc), ratio of the luminosity from OB stars to the total luminosity
  ($L_{OB}/L_{tot}$), total visual extinction measured from the
  surface to the centre of the starburst ($A_v$) and density within
  the hot spots $\rho$ ($cm^{-3}$). Where a range is quoted, e.g.
  $\rho$=100--10000, each of the matched models came up with a value of
  $\rho$  between those limits. Where there is a coma, e.g. 100, 10000,
  each of the matched models had either $\rho$ of 100 or 10000, but not in between.}
\begin{tabular}{|c|c|c|c|c|}
ID &  Nuclear radius & $L_{OB}/L_{tot}$  & $A_v$ & $\rho $ \\
 &  (kpc) & ($\%$) & & ($cm^{-3}$)\\
\hline 
41 &0.35,3 &40,60 &35.9,72 & 100--5000\\
51 &3 & 40,90& 35.9,67.3& 10000\\
55 & 0.35,3 & 40,60& 35.9,72 &100--10000 \\
58 &0.35,1,3  & 40,60&119,144 & 5000,10000\\
67 &1,3  &60 & 35.4,35.9& 10000\\
70 &0.35,1 &40,60 &119,144 & 5000,10000\\
72  &1,3 &40  &2.2--35.9 &1000--10000\\
76  &0.35,1 & 40,60,90&35.9--119&100,10000 \\
77 &1,3 &40,60 & 35.9--72&2500, 10000 \\
78 & 3& 40,60& 35.9,72 &2500,10000  \\
82 & 1,3& 60,90&  18--72& 100--10000 \\
84 & 1,3& 60& 70.7,72 & 10000 \\
85 & 1,3& 40& 2.2,35.9 & 1000,10000\\
88 &0.35,3 & 60,90& 72,120& 10000 \\
93 &0.35,1 & 40,60&35.4--72 & 10000 \\
94 & 1,3& 40&35.4,35.9 & 10000 \\
98 & 1,3&40,60 &4.5--35.4 & 2500,10000  \\
100 &3 & 60,90&2.2--6.7& 100 \\
103 &1,3 & 40,60,90& 2.2--70.7& 100--10000 \\
110 & 0.35,1& 40,60,90& 35.4--144&1000,10000  \\
114 &3 &40 &17.9 & 2500  \\
115 & 3& 40,60& 2.2--72& 1000,10000  \\
120 & 1,3& 40&2.2--9 & 1000,2500  \\
121 & 1,3& 40,60&9--35.9 &  5000\\
122 & 0.35,3& 40,60&35.9--120 &5000,10000  \\
124 & 1,3& 40&17.9,35.9& 7500,10000 \\
125 & 3& 60& 2.2,4.5& 100 \\
126 & 0.35,1& 40,60,90&4.5--17.9 & 100 \\
134 & 1,3&40,60 & 70.7,72& 10000 \\
137 & 1,3& 40& 17.9--72&1000,10000  \\
138 & 1,3& 40& 35.9--72&1000,10000  \\
140 & 1,3&40,60 &9--70.7& 1000--10000 \\
141 & 0.35,1& 40,60&2.2--35.9 & 100--10000 \\
144 & 3&40,60 &2.2 & 100 \\
146 & 1,3&40 & 4.5--35.9&1000,10000  \\
147 &3 &40 &17.9,35.9 & 2500 \\
149 &1,3 & 40&6.7--35.9 & 1000--10000 \\
150 &1,3&60 &17.9,35.9&7500,10000  \\
152 & 1,3& 40,60& 70.7,72& 10000 \\
155 & 0.35,1&40,60,90 & 2.2--6.7& 100 \\
158 & 0.35,3& 40,60&35.9,72 & 2500--10000 \\
172 &0.35 & 40,60,90& 144& 2500\\
\end{tabular}
\end{minipage}
\end{table*}

\subsection{Model Comparison}
The best (lowest $\chi^2$) SED matches for the redshift sample are shown in figures
\ref{fig:badsed}--\ref{fig:goodsed2}. All photometry has been
colour-corrected according to the specifications in the IRAC and MIPS
data handbooks. With the CE01 and DH02 SEDs
we use all available IRAC photometry and the three MIPS bands. In the
case of our empirical SEDs, only the 8--160\,$\mu$m region is taken
into account, because the templates do not extend to short wavelengths. The SK07 models do not treat the near-IR emission in detail, so the fits are approached in two
ways: we fit the 3.6--160\,$\mu$m region (where IRAC photometry is
available) and the 8--160\,$\mu$m region separately. For the latter all
matches with similarly low $\chi^2$ values are kept, in order to increase the accuracy of the
results, since the library contains $\sim$7000 SEDs and limiting the number of
photometric points for the $\chi^2$, dramatically reduces the constraints
on the chosen templates. Furthermore, near-IR photometry is fit with slightly
increased errors in order to take into account the weakness of the
models in that region.

In the mid-IR ($\lambda_{rest}$\,$<$\,40\,$\mu$m) all templates
deliver the emission/absorption lines and continuum slope with the
same consistency; in some cases the 8\,$\mu$m colour-corrected flux
samples a PAH feature (e.g. objects 98, 100, 103, 122,
125). However, there are large discrepancies in the 10\,$\mu$m
silicate absorption depth. For the CE01 and DH02 templates it
seems to vary over a small range; this is not the case for the SK07
SEDs, where it can be 50--100 times below the continuum
(e.g. see objects 58, 70, 76, 82, 110). This is a direct consequence of
SK07 incorporating dust self-absorption in their models, an effect
assigned to increasing quantities of dust at low temperatures
(e.g. Mitchell $\&$ Robinson 1981), resulting in an overall shift
of the far-IR peak to longer wavelengths. Our photometry misses the
silicate feature so we cannot constrain its depth, nevertheless its
varied strength in the models is consistent with observations as it is commonly found in LIRG $\&$ ULIRG spectra (e.g. Hao et
al. 2007, Armus et al. 2007) and its existence is correlated with the
magnitude of extinction.

The differences between the templates are more striking in
the far-IR ($\lambda_{rest}$\,$>$\,40\,$\mu$m). The SK07 models are in good agreement with the
data. In addition, most 3.6--160\,$\mu$m fits result in templates
with similar characteristics to the 8--160\,$\mu$m fits, showing that the far-IR is well
constrained even without the $<$\,6\,$\mu$m region. On the contrary, DH02 and
CE01 often underproduce the 160\,$\mu$m flux, by an amount significantly
higher than the 30 per cent photometric uncertainty and up to a factor
of 10 in extreme cases (e.g. see objects 58,
70, 84, 93, 94, 110, 134, 137, 138, 152, 172 for which the effect is
more exaggerated). Moreover, it is worth noting that most objects for
which the far-IR is completely underestimated by
CE01 and DH02, have a ratio of
$L_{70}/L_{24}>$\,3.4 (table 3), suggesting that a steep mid-IR
continuum and a high far-IR flux are contradictory features in these templates.

\begin{figure*}
\epsfig{file=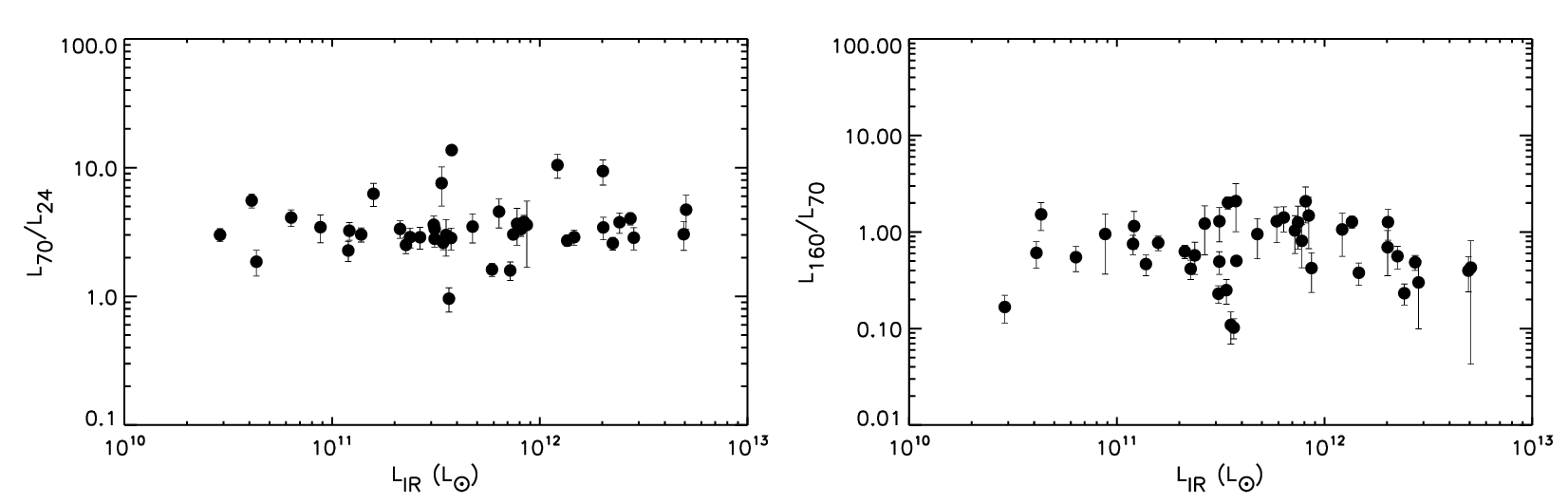,width=15cm}
\caption{The K-corrected, rest frame luminosity ratios $L_{70}/L_{24}$,
  $L_{160}/L_{70}$ as a function of the total infrared luminosity. All
  units are $L_{\odot}$ and the 1$\sigma$ uncertainties are shown.}
\label{fig:lumratios}
\end{figure*}

\begin{figure*}
\epsfig{file=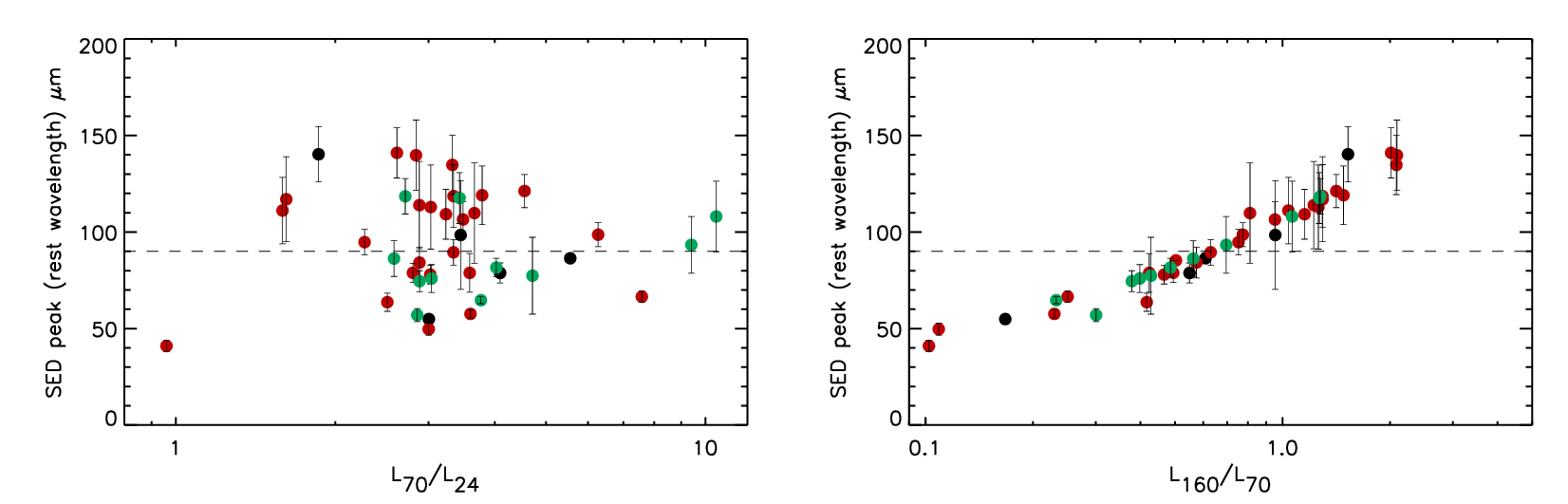,width=15cm}
\caption{The position of the rest frame SED peak (in $\mu$m) as a
  function of the K-corrected, rest frame $L_{70}/L_{24}$ ratio (left panel) and
  $L_{160}/L_{70}$ ratio (right panel). $L_{24}$, $L_{70}$ and $L_{160}$ are
  in units of $L_{\odot}$ and the 1$\sigma$ uncertainties are shown. The colours correspond to the luminosity
  class: black for Starbursts, red for LIRGs and green for ULIRGS.}
\label{fig:lratiopeak}
\end{figure*}

Table 2 shows the parameters generated from the best matched
SK07 templates. The nuclear radius represents the region of total dust
emission from the entire stellar population, with the OB stars only
confined to the inner 350pc. The majority of best fit models are
described by a radius of 3kpc, suggesting
emission from extended regions of cool dust. In terms of visual
extinction, the matches span the whole
available range of $2 < A_v < 144$. Taking into account the
variations of $A_v$ for each object with respect to each template, it is
worth noting that only 5 out of 42 galaxies are matched solely with
$A_v<10$ templates, the remaining
appearing mostly with $A_{v} \sim$ 35 or 70, implying
high obscuration in these systems. Finally, the value of hot spot dust
density, which is directly related to the strength of mid-IR emission appears to have a
wide range in values, but in most objects it reaches the maximum
of $\rho$\,=\,10000\,cm$^{-3}$.

\begin{table*}
\begin{minipage}{126mm}
\caption{The infrared budget and SED characteristics of the redshift sample. Column 1
  classifies the objects according to their luminosity. Columns 2 and
  3 are the Object ID and redshift, columns 4 and 5 are the
  K-corrected, rest-frame
  $L_{70}/L_{24}$ and $L_{160}/L_{70}$ ratios. Columns 6 and 7 quote
  the average value of the total infrared luminosity $L_{IR}$ (in units of
  $\times 10^{10}L_{\odot}$) and its logarithm. The position of the
  far-IR turnover is displayed in column 8. All
  uncertainties are 1$\sigma$.}
\begin{tabular}{|c|c|c|c|c|c|c|}
Type & ID& $L_{70}/L_{24}$ & $L_{160}/L_{70}$ & $L_{IR} \times 10^{10}(L_{\odot})$ &$ log L_{IR}$ & peak (rest, $\mu$m) \\
\hline 
&      51&       1.86&       1.52&       4.32$\pm$      0.23&       10.64&       140$\pm$
       14\\
&     155&       3.00&      0.17&       2.89$\pm$      0.2&       10.46&       55$\pm$
       2\\
SBs&     125&       4.10&      0.55&       6.35$\pm$      0.72&       10.81&       79$\pm$
       5\\
&     144&       5.56&      0.61&       4.10$\pm$      0.49&       10.61&       86$\pm$
       2\\
\hline
SB/LIRG&     103&       3.45&      0.95&       8.79$\pm$       1.31&       10.94&       98$\pm$
       28\\
\hline

&     141 &      0.96 &      0.10 &       36.54$\pm$       1.66 &       11.56 &       41$\pm$       3\\
&     121 &       1.59 &       1.04 &       72.07$\pm$       8.26 &       11.86 &       111$\pm$       17\\
&     150 &       1.62 &       1.29&       58.88$\pm$       9.01 &       11.77 &       117$\pm$       22\\
&      98 &       2.27 &      0.75 &       11.97$\pm$      1.0 &       11.08 &       95$\pm$       7\\
&     100 &       2.51 &      0.42 &       22.68$\pm$       1.48 &       11.36 &       64$\pm$       5\\
&      78 &       2.62 &       2.01 &       34.27$\pm$       1.03 &       11.53 &       141$\pm$       13\\
&     120 &       2.80 &      0.49 &       31.15$\pm$       2.84 &       11.49 &       78$\pm$       5\\
&     152 &       2.84 &       2.09 &       37.41$\pm$       8.58 &       11.57 &       140$\pm$       18\\
&     115 &       2.88 &       1.22 &       26.50$\pm$       7.93 &       11.42 &       114$\pm$       23\\
&      85 &       2.88 &      0.57 &       23.73$\pm$       2.32 &       11.37 &       84$\pm$       8\\
&      76 &       3.00 &      0.11 &       35.41$\pm$       3.65 &       11.55 &       50$\pm$       3\\
&      72 &       3.02 &      0.47&       13.81$\pm$      0.87 &       11.14 &       78$\pm$       5\\
&     147 &       3.03 &       1.26 &       74.57$\pm$       16.67 &       11.87 &       113$\pm$       22\\
 LIRGs &   122 &       3.24 &       1.15 &       12.11$\pm$       2.31 &       11.08 &       109$\pm$       13\\
&      77 &       3.33 &       2.08&       81.11$\pm$       16.14 &       11.91 &       135$\pm$       15\\
&     146 &       3.34 &      0.63 &       21.24$\pm$      0.98 &       11.33 &       90$\pm$       7\\
&     140 &       3.35 &       1.29 &       31.18$\pm$       4.74 &       11.49 &       119$\pm$       16\\
&     149 &       3.48 &      0.95 &       47.50$\pm$       7.27 &       11.68 &       107$\pm$       9\\
&      55 &       3.59 &      0.42 &       86.61$\pm$       3.77 &       11.94 &       79$\pm$       10\\
&     126 &       3.60 &      0.23 &       30.87$\pm$       1.90 &       11.49 &       58$\pm$       2\\
&      41 &       3.66 &      0.81 &       77.76$\pm$       8.38 &       11.89 &       110$\pm$       26\\
&     137 &       4.55 &       1.41 &       63.65$\pm$       4.0 &       11.80 &       121$\pm$       8\\
&     138 &       6.27 &      0.77 &       15.82$\pm$      0.59 &       11.20 &       99$\pm$       6\\
&     110 &       7.58 &      0.25 &       33.68$\pm$       3.37&       11.53&       67$\pm$       3\\
&     172 &       13.69 &      0.50 &       37.65$\pm$      0.12 &
11.58 &       85$\pm$       0\\
\hline
(U)LIRG &     134 &       3.79 &       1.48 &       83.94$\pm$       21.39 &       11.92 &       119$\pm$       15\\
\hline
&      67&       2.58&      0.56&       224.71$\pm$       2.74&       12.35&       86$\pm$
       9\\
&     114&       2.71&       1.27&       135.64$\pm$       1.54&       12.13&       119$\pm$
       9\\
&      82&       2.86&      0.30&       283.69$\pm$       39.97&       12.45&       57$\pm$
       3\\
&     124&       2.89&      0.38&       146.39$\pm$       8.87&       12.17&       75$\pm$
       5\\
ULIRGs&      93&       3.04&      0.40&       493.36$\pm$       55.53&       12.69&       76$\pm$
       7\\
&      84&       3.43&       1.27&       202.19$\pm$       8.25&       12.31&       118$\pm$
       13\\
&      88&       3.77&      0.23&       242.27$\pm$       15.29&       12.38&       65$\pm$
       2\\
&      94&       4.03&      0.49&       273.44$\pm$       2.96&       12.44&       82$\pm$
       5\\
&     158&       4.71&      0.43&       506.70$\pm$       111.65&       12.70&       77$\pm$
       20\\
&      70&       9.40&      0.69&       201.46$\pm$       27.19&       12.30&       93$\pm$
       15\\
&      58&       10.48&       1.06&       121.73$\pm$       16.74&       12.08&       108$\pm$
       18\\

\hline 
\end{tabular}
\end{minipage}
\end{table*}

\subsection{Summary}

For the 8--160\,$\mu$m region, the SK07 models provide the best matches
to our sources' infrared SEDs. The fact that the CE01 and DH02 templates in most cases severely underestimate the far-IR, leads us to two
conclusions. Firstly, the notion of a single radiating source is an
over-simplification of a real system, which is more likely to be a
complex and clumpy 3-D configuration with a varying temperature
gradient. Secondly, we propose that these high-z objects have higher
obscuration and additional far-IR emission from a cold dust component that is not commonly seen in
local LIRGs/ULIRGs, on which the CE01 and DH02 templates are
based. Such excess far-IR flux would be a direct consequence of a large emitting dust region, where dust
in the immediate vicinity of young stars (and/or an AGN), would be
responsible for greatly attenuating the radiation field and hence
shielding the outer dust layers from direct UV illumination, resulting
in lower equilibrium temperatures. This scenario is well-expressed by the SK07
models; by including the treatment of hot spots around OB stars,
the average dust temperature is lowered, shifting the far-IR peak to
longer wavelengths, without a reduction in the mid-IR flux, which,
according to their framework, is only linked to the dust density in
the hot spot. In contrast, the CE01 and DH02 templates, which have been
found to work well with local sources, fail in this case, because
their formulation does not allow for the possibility of excess far-IR
flux independent from the mid-IR. These observations are further supported by our empirical
SED fits, which were included for the purpose
of comparing with real local galaxy data (see
section 3.2). The majority show considerable
discrepancies, failing to reproduce the far-IR SED shape (e.g. objects
77, 93, 110, 122), indicating that the greybody temperatures of 30--50\,K are too high and that dust
potentially colder than 30\,K is present in most systems.

In view of these results, our subsequent
 calculations consider only the SK07 fits to the
 8--160\,$\mu$m spectral region. The fact that
the near-IR part of the SED is not well-represented, is not important since we are
only concerned with pure dust emission in the 8--1000\,$\mu$m
 range. For each object, we take into account
all SK07 matches (figures \ref{fig:goodsed} and \ref{fig:goodsed2} and
table 2), to establish a valid error
margin in cases where the 160\,$\mu$m detection could either reside
on the peak or form part of the rising continuum (e.g. see objects
 115, 147, 152).

\section{SED characterisation}

\subsection{The Infrared Energy Budget}
With the aid of the best-matched SK07 templates, we perform
K-corrections and estimate the monochromatic rest frame
luminosities at 24, 70 and 160\,$\mu$m. In addition, by integrating the SEDs between
8--1000\,$\mu$m, we arrive at an array of values for the total infrared
luminosity ($L_{IR}$) for each
object, the average of which is then used as a final estimate.
 The sources in the redshift sample are categorised into
luminosity classes, following the
definition in Sanders $\&$ Mirabel (1996), revealing that it consists
of 12 per cent Starbursts, 62 per cent LIRGs and 26
per cent ULIRGs. Despite the uncertainties on $L_{IR}$ being
relatively large and stemming from the lack of photometry longward of 160\,$\mu$m,
only two objects move to a different luminosity class on account of
this (table 3). 

\subsection{The mid-IR continuum slope and SED peak}
Conventional methods of characterising infrared galaxies include associating
flux ratios between various bands to physical
processes responsible for emission in the infrared (e.g. Xu et al. 2001;
Egami et al. 2004; Lutz et al. 2005; Verma et al. 2005). In
particular, the $f_{25}/f_{60}$ parameter has been applied
extensively in \emph{IRAS} studies of such systems, to quantify
relative AGN/starburst contributions with an indicative value of $f_{25}/f_{60}$\,$>$\,0.17 for AGN-dominated sources (e.g. de Grijp et
al. 1985, Miley et al. 1985, Soifer et al. 1989). This `warm/cold'
terminology is based on the implication that dust heated by an AGN
would reach higher temperatures and hence lower the continuum slope between 25 and
60\,$\mu$m (or 24 and 70\,$\mu$m in \emph{Spitzer}'s case) (e.g. Farrah et
al. 2005; Verma et al. 2005, Frayer et al. 2006). Calculating
the 25/60\,$\mu$m continuum slope using the model SEDs, reveals that
for $\sim$25 per cent of our sample the classical IRAS warm source
criterion of $f_{25}/f_{60}$\,$>$\,0.17 applies. However, the majority of
these objects peak at $\lambda>$\,100\,$\mu$m, indicating that the
flatter behaviour between 25 and 60\,$\mu$m is due to a shift in
infrared emission with elevated $\lambda>$\,90\,$\mu$m flux and reduced 60\,$\mu$m flux. Furthermore, from figure
\ref{fig:lumratios}, it is clear that the continuum slope throughout
the SED is completely independent from the total infrared
luminosity. Consequently, if it is true that the contribution of
AGN-dominated objects increases as a function of IR luminosity
(e.g. Brand et al. 2006), then the slope of the mid-IR continuum and
hence the $L_{70}/L_{24}$ ratio, fail to
quantify it. In addition, because of heavy obscuration,
emission from cool dust seems to be altering the mid-IR
continuum slope solely due to increased far-IR
emission. Various authors have
supported this idea; Chakrabarti et al. (2007) have modelled infrared
SEDs with varied starburst/AGN contributions, suggesting that the warm/cold classification of LIRGs/ULIRGs would be independent of the energy source, if sufficiently
obscured so that all the far-IR emission comes from large grains in
equilibrium. Similar observations (e.g.  Lutz et al. 1996, Klaas et
al. 1997) have put forward the notion that elevated
optical extinction can be responsible for a steep rise of the mid-IR continuum and hence can displace the ratio of the `cold -warm' components, resulting in an energy output mainly carried by the cold component. 
  
To further investigate the properties of this sample, the wavelength at which the SEDs turn over into the
Rayleigh-Jean regime is estimated (table 3, column 8). We find that a
significant fraction of objects (40-60 per cent within 1$\sigma$) peak at wavelengths longward of
90\,$\mu$m independent of luminosity class. Despite
the large uncertainties, there is clearly a strong correlation with respect
to $L_{160}/L_{70}$, not surprising, since 70 and 160\,$\mu$m sample
either side of the peak and their ratio is directly related to the
far-IR turnover (figure \ref{fig:lratiopeak}, right panel). The same does not apply for the $L_{70}/L_{24}$
ratio, however, which shows no such dependence (figure \ref{fig:lratiopeak}, left panel). 
 
Figures \ref{fig:lumratios} and \ref{fig:lratiopeak} display some
important properties of the sample, consistent with conclusions
from section 3. Generally, a simple black body spectrum displays a
temperature dependence through the position of the emission peak, the continuum
slope and the total luminosity. Consequently, if the
far-IR part of an SED were well-represented by a [modified] black body, for a sample of galaxies of
varying dust temperatures this would manifest itself
as: a) a positive correlation between mid-IR continuum slope and total infrared
luminosity (as they both increase with temperature) and b) a negative
correlation between mid-IR continuum slope and position of SED peak (as the former increases and the
latter decreases with temperature). However, if an extra cold dust
component is present, as we believe the case to be for the 70\,$\mu$m
sample, a \emph{single} temperature black body is not a
sufficiently accurate representation of the far-IR and these
correlations would not apply. This is what is observed in figures
\ref{fig:lumratios} and \ref{fig:lratiopeak}.

\begin{figure}
\epsfig{file=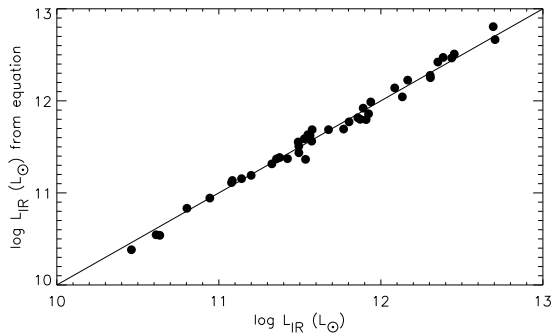,width=8cm}
\caption{The relation between $L_{IR}$ calculated using equation 1 (y-axis) and
  $L_{IR}$ calculated by integrating under the SEDs (x-axis). The
  solid line has a gradient of 1. The r.m.s. scatter off the line is 0.06
  dex.}
\label{fig:formulagood}
\end{figure}

\section{Equations for estimating the Total Infrared Luminosity}

Determining the infrared energy budget is crucial for
estimating the contribution to the infrared and submillimetre
backgrounds and plays an important role in establishing a value for
the infrared element of the cosmic star formation history. If the
available photometric data is not sufficient and hence unable to explicitly define
these contributions, a first order approximation of the infrared
energy budget can be attained through equations relating the total
infrared luminosity to the monochromatic luminosity in one or more bands.
In this section, we develop equations that can be used to estimate
$L_{IR}$, using 1 or more MIPS bands. However, note that all have been
derived using the typical form of starburst-type SEDs and have not
been tested on power-law SEDs of AGN-dominated sources.

\subsection{Estimating $L_{IR}$ with all 3 MIPS bands}
 As the SED of an infrared-luminous galaxy has a distinct shape, its
 bolometric output can be approximated by summing contributions from
 four main sections: $<$\,24\,$\mu$m, 24--70\,$\mu$m, 70--160\,$\mu$m and $>$\,160\,$\mu$m. In a linear plot of $f_{\nu}$ versus $\nu$, flux shortward of 24\,$\mu$m and
 longward of 160\,$\mu$m is not as significant as the flux
 within that range. Therefore the $<$\,24\,$\mu$m and
 $>$\,160\,$\mu$m sections can be approximated by
 triangles of heights $L_{24}$ and $L_{160}$ and the sections
 within 24--160\,$\mu$m with trapezia of heights of $L_{24}$, $L_{70}$ and
 $L_{70}$, $L_{160}$. Summing these contributions and empirically
 accounting for the systematic offset in this approximation, we derive an
 equation to calculate the total infrared luminosity ($L_{IR}$) in the range
 8-1000\,$\mu$m:
\begin{equation}
L_{IR}=4.63\times 10^{-15}\times(8.3L_{24}+2.7L_{70}+L_{160})\times (1+z).
\label{eqn:lum1}
\end{equation}
The coefficients represent the width ($\Delta \nu$) of the bases of the sections
 in units of Hz, $L_{IR}$ is in units of $L_{\odot}$ and $L_{24}$,
 $L_{70}$ and $L_{160}$
are the monochromatic luminosity densities (in units of
 W/Hz) directly derived from the observed flux densities, without
 applying K-corrections. The factor (1+z) adjusts the bases $\Delta \nu$
 of the SED sections, as their width increases with redshift. 

\begin{figure}
\epsfig{file=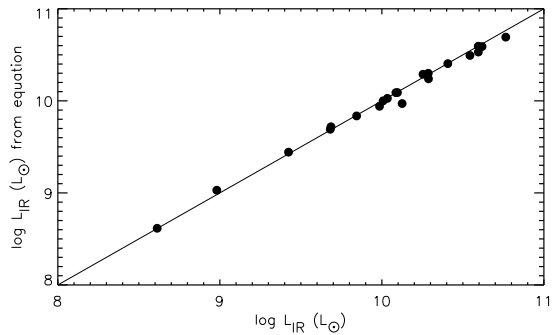,width=8cm}
\caption{The relation between $L_{IR}$ calculated using equation 1
  (y-axis) and $L_{IR}$ calculated by integrating under the SEDs
  (x-axis) for 21 SINGS galaxies. The
  solid line has a gradient of 1. The r.m.s. scatter off the line is 0.05
  dex.}
\label{fig:sings}
\end{figure}

The validity of
 this method is demonstrated by comparing
the value of $L_{IR}$ derived by using the equation (y-axis) and the
average value of $L_{IR}$ derived by integrating under the best-matched
model SEDs (x-axis) (figure \ref{fig:formulagood}). There is very good
 agreement, with an r.m.s. deviation of 0.06
 dex from the true $L_{IR}$. We also test it on a different galaxy sample, by calculating the total infrared luminosity of 21
local infrared star-forming/starburst galaxies using MIPS photometry
from the \textit{Spitzer} Infrared Nearby Galaxy Survey
(SINGS) (Dale at al. 2005, hereafter D05) and estimated
distances quoted in Sanders et al. (2003). We fit the DH02 models on
the 24, 70 and 160\,$\mu$m photometry (like in D05) and integrate under
the 8-1000\,$\mu$m region of the best-fit curve, comparing the result with
$L_{IR}$ from equation 1. The agreement is very good, with an r.m.s. scatter of
0.05 (figure \ref{fig:sings}), confirming that this equation is not
 specific to high-z LIRGs/ULIRGs, but also applicable to normal
 infrared star-forming galaxies in the local universe.

\begin{figure}
\epsfig{file=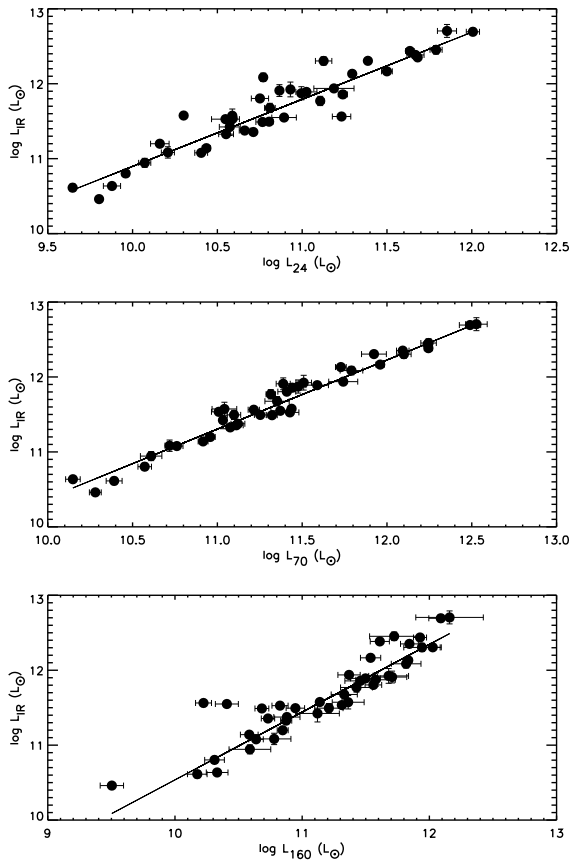,width=8cm}
\caption{The monochromatic rest frame luminosity at 24, 70 and
  160\,$\mu$m as a function of the total infrared luminosity for the
  redshift sample. The 1$\sigma$ uncertainties are shown. The r.m.s. scatter is 0.18, 0.11,
and 0.23 dex respectively.}
\label{fig:mono}
\end{figure}

\subsection{Estimating $L_{IR}$ with 1 MIPS band}
Naturally, a more luminous galaxy will also display higher luminosity
at the various infrared photometric bands (e.g. Elbaz et
al. 2002, Takeuchi et al. 2005). This would manifest itself as a
strong positive correlation between $L_{IR}$ and K-corrected
rest-frame monochromatic
luminosity at 24, 70 and 160\,$\mu$m (figure \ref{fig:mono}). As a
result, it could be used instead of equation 1, for estimating $L_{IR}$, in cases where only one photometric data point
from MIPS is available. For each plot in figure \ref{fig:mono}, we compute the least squares
best fit line through the data, equations 2--4:

\begin{equation}
\rm log \,L_{IR}=1.95 + 0.89\rm \;log\,L_{24},
\end{equation}

\begin{equation}
\rm log \,L_{IR}=1.16 + 0.92\rm \;log\,L_{70}, 
\end{equation}

\begin{equation}
 \rm log\,L_{IR}=1.49 + 0.9\rm \;log\,L_{160},
\end{equation}
 All luminosities are in units of $L_{\odot}$ and $L_{24}$, $L_{70}$ and
$L_{160}$ are in the rest frame. The r.m.s. offsets from the best line
fits are 0.18, 0.11, 0.23 dex for equations 2, 3 and 4 respectively. This intrinsic dispersion most likely stems from the
variation in the properties of each galaxy, the high
photometric uncertainties as well as the range of models used to
calculate each luminosity value; nevertheless, they are convenient in
cases where the luminosity of a galaxy is required only within an
order of magnitude or class.

\section{Summary and Conclusion}

We have examined a population of 70\,$\mu$m selected sources, in the region
of the Extended Groth Strip, focusing this study on the infrared
properties of 43 objects with
available spectroscopic redshifts (0.1\,$<$\,z\,$<$\,1.2). In the last
section, we derived a set of equations in order to estimate the total
IR luminosity with at least one MIPS band, applicable to objects with starburst-type SEDs. We would like to emphasise that our selection criteria did not
photometrically bias the redshift sample with respect to the full
sample or predispose our study to a particular type of galaxy. As
this is a flux limited survey (see section 2), the higher redshift objects are also more
luminous in the infrared and display lower optical flux. This implies that the sources with no
available spectroscopic redshift are also in the LIRG/ULIRG regime. In addition, the MIPS colours (figure
\ref{fig:mipscolours}) and SEDs (figure
\ref{fig:seds}) unanimously show that all objects are
photometrically similar in the mid and far-IR and their properties
diverge only in the optical and near-IR, not relevant for the
work described here. 

By fitting 4 libraries of SED templates on our infrared photometry, we
have examined the properties of the sources and estimated the infrared energy budget, revealing 12 per cent starbursts, 62 per cent LIRGs
and 26 per cent ULIRGs. Evaluation of
the fits showed that the SK07 models perform better at reproducing the
far-IR SED; their formulation adopts a
two-component configuration of dust heated in the immediate locality
of massive young stars and cirrus-heated dust, essentially decoupling
mid and far infrared emission. As a result, since the total infrared
luminosity is not directly related to the position of the SED peak or
the steepness of the mid-IR continuum slope, with the SK07 templates
it is possible to represent objects with strong mid-IR emission and an
additional cold component. CE01 and DH02 employ a scenario with a
single radiating central source where temperature varies as a function
of distance from the centre. Although this has worked well with local
infrared galaxies, it completely underestimates our sources'
160\,$\mu$m flux, with a discrepancy of at least a factor of 2 and up
to 10 for extreme cases, since it does not allow for types of systems
with elevated far-IR emission. Our empirical local galaxy templates,
also mostly underestimate the far-IR region, indicative of the fact that an average
dust temperature of 30-50\,K is too high.

Considering all above
points, we propose that this deep
70\,$\mu$m survey has probed high-z LIRGs and ULIRGs, which are unlike local
counterparts: heavy
obscuration and large amounts of cold dust potentially lower than 30\,K, appear as a \emph{far-infrared excess}
component in the SEDs, causing a significant fraction to peak at
$\lambda>$\,90\,$\mu$m and steepening the continuum slope. The existence
of such objects has also been put forward by other authors; e.g. Marcillac et
al. (2006) have suggested that some high-z infrared-luminous objects display far-IR properties
divergent from those of local galaxies because of an additional cold
dust component, even if the strength of mid-IR emission is the same.
This has further implications: in such systems, the
continuum slope cannot be representative of various
galaxy characteristics such as dust temperature,
total infrared luminosity or the nature of the
central energy source. Figures \ref{fig:lumratios}
and \ref{fig:lratiopeak} demonstrate this: the mid-IR continuum slope (quantified
by the $L_{70}/L_{24}$ ratio) satisfies the IRAS warm source criterion
of $f_{25}/f_{60}$ $>0.17$ for sources with reduced 60\,$\mu$m
emission and increased $\lambda >$\,90\,$\mu$m flux, as opposed to
elevated 25\,$\mu$m emission and it is completely
dissociated from the position of the SED turnover and total infrared luminosity. As this is inconsistent with the representation of far-IR emission as
a single temperature black body, it is in line with our earlier
suggestion of the presence of an additional cold emissive
component.

It is worth noting that this 70\,$\mu$m sample is one of
the first far-IR-selected samples reaching such low flux-density
limits and additional similar surveys will
be of great importance in highlighting the differences (if any)
between local and high redshift infrared populations. The next infrared
observatory, \emph{Herschel}, will greatly benefit this work,
extending to the sub-mm part of the spectrum and, hence, enabling a
complete census of high-z infrared galaxy SEDs. Finally, we would like
to mention that in order to confirm our conclusions we have planned a
follow-up of the work described in this paper with
submillimetre observations using the Submillimetre Common-User Bolometer Array (SCUBA 2).

\section*{Acknowledgments}
M.S. has received support from the University of Oxford, department of Astrophysics. This work is based on observations made with the Spitzer Space Telescope, operated by the Jet Propulsion Laboratory, California Institute of Technology, under NASA contract 1407. We thank Pat Roche, Aprajita Verma, Steve Rawlings and Tom Mauch for useful discussions and comments.

\label{lastpage}

\end{document}